\documentclass[sigconf]{acmart}

\setcopyright{iw3c2w3}
\copyrightyear{2021}
\acmYear{2021}
\acmConference[WWW '21]{Proceedings of the Web Conference 2021}{April 19--23, 2021}{Ljubljana, Slovenia}
\acmBooktitle{Proceedings of the Web Conference 2021 (WWW '21), April 19--23, 2021, Ljubljana, Slovenia}
\acmPrice{}
\acmDOI{10.1145/3442381.3449940}
\acmISBN{978-1-4503-8312-7/21/04}

\AtBeginDocument{%
  \providecommand\BibTeX{{%
    \normalfont B\kern-0.5em{\scshape i\kern-0.25em b}\kern-0.8em\TeX}}}

\usepackage{bm}
\usepackage{multirow}
\usepackage{balance}

\begin{document}
\fancyhead{}
\title{Dual Side Deep Context-aware Modulation for Social Recommendation}  

\author{Bairan Fu}
\email{fubr@smail.nju.edu.cn}
\affiliation{%
  \institution{Nanjing University}
  \city{Nanjing}
  \country{China}
}

\author{Wenming Zhang}
\email{zhangwm@smail.nju.edu.cn}
\affiliation{%
  \institution{Nanjing University}
  \city{Nanjing}
  \country{China}
}

\author{Guangneng Hu}
\email{njuhgn@gmail.com}
\affiliation{%
  \institution{Hong Kong University of Science and Technology}
  \city{Hong Kong}
  \country{China}
}

\author{Xinyu Dai}
\authornote{Corresponding author.}
\email{daixinyu@nju.edu.cn}
\affiliation{%
  \institution{Nanjing University}
  \city{Nanjing}
  \country{China}
}

\author{Shujian Huang}
\email{huangsj@nju.edu.cn}
\affiliation{%
  \institution{Nanjing University}
  \city{Nanjing}
  \country{China}
}

\author{Jiajun Chen}
\email{chenjj@nju.edu.cn}
\affiliation{%
  \institution{Nanjing University}
  \city{Nanjing}
  \country{China}
}

\renewcommand{\shortauthors}{Fu, et al.}

\begin{abstract}

Social recommendation is effective in improving the recommendation performance by leveraging social relations from online social networking platforms. Social relations among users provide friends' information for modeling users' interest in candidate items and help items expose to potential consumers (i.e., item attraction). However, there are two issues haven't been well-studied: Firstly, for the user interests, existing methods typically aggregate friends' information contextualized on the candidate item only, and this shallow context-aware aggregation makes them suffer from the limited friends' information. Secondly, for the item attraction, if the item's past consumers are the friends of or have a similar consumption habit to the targeted user, the item may be more attractive to the targeted user, but most existing methods neglect the relation enhanced context-aware item attraction.

To address the above issues, we proposed DICER (\textit{\textbf{D}ual s\textbf{I}de deep \textbf{C}ontext-awar\textbf{E} modulation for social \textbf{R}ecommendation}). Specifically, we first proposed a novel graph neural network to model the social relation and collaborative relation, and on top of high-order relations, a dual side deep context-aware modulation is introduced to capture the friends' information and item attraction. Empirical results on two real-world datasets show the effectiveness of the proposed model and further experiments are conducted to help understand how the dual context-aware modulation works.

\end{abstract}

\ccsdesc[500]{Information systems~Social recommendation}
\ccsdesc[500]{Information systems~Recommender systems}

\keywords{recommender systems, social recommendation, graph neural networks, context-aware recommendation}


\maketitle

\section{Introduction}

With the rapid development of the Internet, information overload is becoming increasingly challenging in providing personalized information for users. Recommender systems designed to filter information and provide personalized recommendations play a vital role in various web services nowadays. Widely used recommendation methods are based on Collaborative Filtering (CF) techniques~\cite{koren2009matrix,koren2008factorization, hu2008collaborative, rendle2009bpr}, which mainly make use of the user-item interaction history, e.g., ratings, clickings. Moreover, along with the increasing popularity of social networking platforms, social recommendation, which incorporates social relations into recommender systems, has been developed and shows promising potential to improve the recommendation performance. Social relations among users can provide friends' information for modeling user preference better and also provide more possible perspectives for items' exposure to relevant users~\cite{ren2017social}.

\begin{figure}[tbp]
	\centering
	\includegraphics[width=1\linewidth]{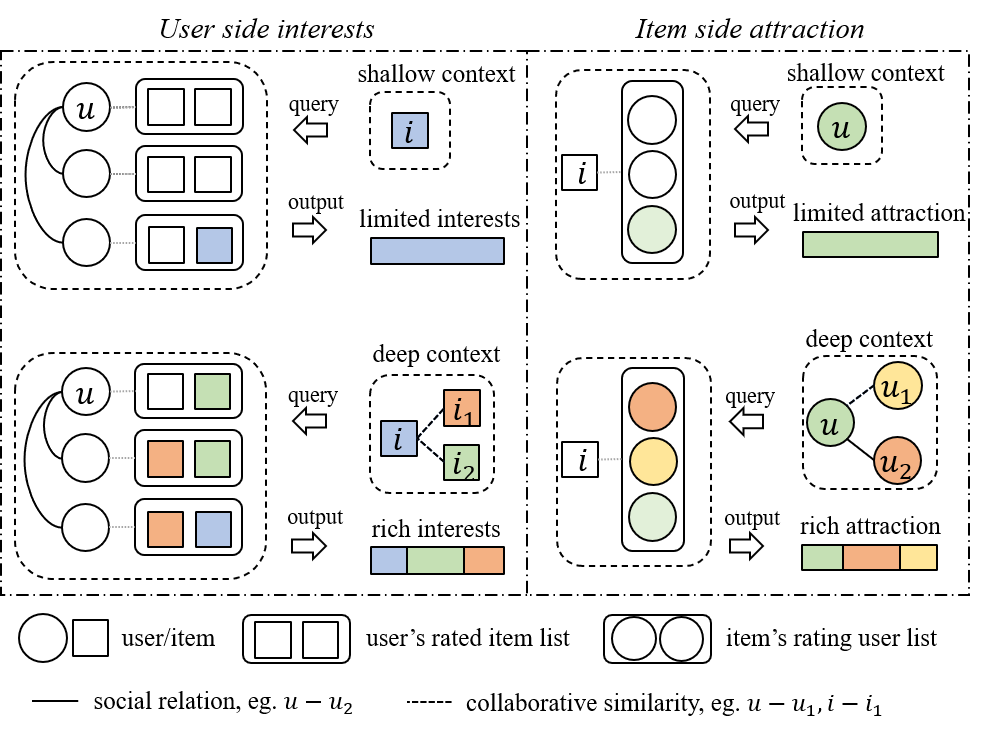}
    \caption{Illustration of dual side deep context. The left part is user side interest while the right part is item side attraction. Deep context have been shown to be helpful to model users' rich interests and items' rich attraction. }
    \label{fig:motivation}
\end{figure}

To incorporate social relations into the recommender system, existing social-aware recommendation methods make several attempts in various ways. Typical matrix factorization methods~\cite{jamali2010matrix, ma2011recommender, jiang2014scalable, guo2016a} assume that users who have social relations may have a similar preference. Thus, these methods use social relations as a kind of social regularization to restrain the user embedding learning process. Moreover, some other methods~\cite{guo2015trustsvd, chen2019social} consider that connected people would influence each other based on the social influence theory. These methods incorporate friends' opinion on candidate items to model the user's preference. For example, TrustSVD~\cite{guo2015trustsvd} incorporates the friends' decisions to model the users' preference and SAMN~\cite{chen2019social} design a two-level attention mechanism to model the friends' influence. However, these methods only consider the first-order local neighbors' information and neglect the helpful information from distant neighbors.

Recently, deep neural networks for graph data, which are known as Graph Neural Networks (GNNs), have shown an effective performance and have experienced rapid development. The core idea of GNNs is about the propagation and aggregation of information from neighbor nodes~\cite{kipf2017semi, bruna2014spectral,NIPS2016_6081}, which naturally accord with social recommendations. Several recent works that utilize GNNs for social recommendations~\cite{fan2019graph, wu2019a,wu2019dual, wu2020diffnetplus} have been proposed. For example, the DiffNet++ method~\cite{wu2020diffnetplus} developed a GNN based model to simulate both the social influence and user interest diffusion process. The DANSER model~\cite{wu2019dual} developed a dual-graph attention network to model the two-fold social effects collaboratively.

Although the aforementioned social recommendation methods have shown performance improvement, they do not fully take advantage of social network information in a deep way. First, when modeling user side interest, most methods consider the friends' information equally without considering the specific recommendation context. This results in a lot of noisy information are aggregated from friends' information. And some other methods consider the candidate item as a context to model the context-aware friends and user's interest (as shown in the top-left part of Figure ~\ref{fig:motivation}). However, considering the candidate item as a (shallow) context only leads to interest information biased to some extend, and thus limited user and friends' interest can be obtained. In fact, when modeling user side interests from user's and friends' historical interacted items, not only the information related to the candidate item can help, the information related to candidate's similar items can also reflect the user's interests (as shown in the bottom-left part of Figure ~\ref{fig:motivation}), especially when the candidate item has only limited interaction history. 

Furthermore, the related information from the candidate item's interaction history to the targeted user also reflects the item's attraction to the users. Few social recommendation work~\cite{wu2019dual} models the item attraction based on the targeted user. As shown in the top-right part of Figure ~\ref{fig:motivation}, considering the targeted user only is also a shallow context-aware method and results in poor item attraction being modeled. In fact, if the item's past consumers is a friend of the targeted user or have the similar consumption habit to the target user, the item may be more attractive to the users, as shown in the bottom-right part of Figure ~\ref{fig:motivation}. Hence, it is obvious that considering the social relation and similarity relations among users helps to extract more useful information from the candidate item's interaction history.

While it is of great potential to leverage social relations and similarity relations in mining interaction history information for the recommendation, there are still several significant challenges. First, the high-order social relations and similarity relations are complex and it is not easy to extract the most related information for modeling user preference and item attribute. Second, it is not trivial to extract the user interest and item attraction from the interaction history based on the high-order relation enhanced context, as there may be much noisy information.

To tackle the challenges mentioned above, in this paper, we proposed DICER (\textit{\textbf{D}ual s\textbf{I}de deep \textbf{C}ontext-awar\textbf{E} modulation for social \textbf{R}ecommendation}) which incorporating the high-order neighbor information to model the enhanced user preference and item attribute, and extracting the most related information from the interaction history based on the graph enhanced deep context. 

Our contribution in this paper can be summarized as follows:
\begin{itemize}
    \item We propose a novel neural model for the social recommendation, which utilizes relation-aware Graph Neural Network to use multi-relation and high-order neighbor information effectively. A Deep Context-aware Modulation is introduced to model user side interests and item side attraction.
    \item To the best of our knowledge, we are the first to consider the social relation and collaborative similarity among users as deep context to model the item attraction representations. In duality, we also consider the collaborative similarity among items as deep context to model the user and friends' interest representations.
    \item Experiments on two real-world benchmark datasets are conducted to demonstrate the effectiveness of our model. We show that our model consistently outperforms the state-of-the-art models.
\end{itemize}

The remainder of this paper is organized as follows. We introduce the proposed model in Section 3. In Section 4, we conduct experiments on two real-work datasets to demonstrate the effectiveness of the proposed model. In Section 5, we review related works. Finally, we conclude with future directions in Section 6.

\section{Preliminary}

We first introduce the and notations used throughout the paper and then describe the problem formulation.

\subsection{Notations}

We define \( U (|U|=M)\) and \( I (|I|=N) \) as the sets of users and items respectively. We use \(u\),\(v\) to index users, and \(i\), \(j\) to index items. We consider a user-item interaction matrix \(\mathbf{R} = [r_{u,i}]_{M\times N} \in \{0,1\} \) indicating whether \(u\) has interacted (e.g., purchased, clicked) on item \(i\). We use \(R_I(u)\) and \(R_U(i)\) to respectively denote the set of items rated by user \(u\) and the set of users who have rated item \(i\). 

We consider the user-user social network as a graph \(G^{S}_U = (V_U, E^{S}_U)\) where \(V_U\) is the set of users and \(E^{S}_U\) is the set of edges that connect two users. We use \(F_U(u)\) to denote the set of nodes adjacent to \(u\) in \(G^{S}_U\). We define \(\mathbf{P} = \{\mathbf{p}_u\}_{D\times M}\), where \(D\) is the embedding dimension and \(\mathbf{p}_u\) denotes the embedding vector for user \(u\); similarly, \( \mathbf{Q} = \{\mathbf{q}_i\}_{D\times N} \), where \(\mathbf{q}_i\) denotes the embedding vector for item \(i\). The mathematical notations used in this paper are summarized in Table \ref{tab:notations}.

\subsection{Problem Formulation}

The social recommendation problem is defined as~\cite{tang2013social, wu2019dual}: given the user-item interactions \(\mathbf{R}\) and the user-user social network \(G^{S}_U\), the goal is to predict the unobserved interactions in \(\mathbf{R}\), i.e., the probability of a targeted user \(u\) clicking or purchasing an unobserved candidate item \(i\).

\subsection{Background}

\textbf{Graph Neural Network}. Graph Neural Networks have been proposed to learn the graph topological structure and nodes' feature information. And the most representative method is Graph Convolutional Networks (GCN)~\cite{kipf2017semi}, which learns representation for a node by aggregating its neighbors' representation iteratively. GCN treats the neighbor information equally without considering the different importance of the neighbors. In contrast, Graph Attention Networks (GAT)~\cite{velivckovic2017graph} specifying different weights to different neighbors, which helps the model learn more related information from neighbors. Moreover, Neural Graph Collaborative Filtering (NGCF)~\cite{wang2019neural} proposed an effective message propagation approach to aggregate the more similar information from neighbors, which enable the model to model the high-order relation information and filter the noisy neighbor effectively.

\begin{table}

  \caption{Summary of notations}
  \label{tab:notations}
  \resizebox{\linewidth}{!}{
  \begin{tabular}{c|c}
    \hline
    \textbf{Symbols} & \textbf{Definitions and Descriptions} \\
    \hline
    \(\mathbf{U}\) & set of users \\
    \(\mathbf{I}\) & set of items \\
    \(\mathbf{R}\) & user-item interactions \\
    \(G^{S}_U\) &  user-user social network\\
    \(G^{R}_U\) &  user-user collaborative similarity network\\
    \(G^{R}_I\) &  item-item collaborative similarity network\\
    \(R_I(u)\) & the set of items rated by user \(u\)\\
    \(R_U(i)\) & the set of users who have rated item \(i\) \\
    \(F_U(u)\) & the set of nodes adjacent to \(u\) in \(G^{S}_U\) \\
    \(N_U(u)\) & the set of nodes adjacent to \(u\) in \(G^{R}_U\) \\
    \(N_I(i)\) & the set of nodes adjacent to \(i\) in \(G^{R}_i\) \\
    \(\mathbf{p_u}\) & the embedding of user \(u\) \\
    \(\mathbf{q_i}\) & the embedding of item \(i\) \\
    \(h^{S,l}_u\) & the \(l\)-th layer representation of user \(u\) in \(G^{S}_U\) \\
    \(h^{R,l}_u\) & the \(l\)-th layer representation of user \(u\) in \(G^{R}_U\) \\
    \(z^{l}_i\) & the \(l\)-th layer representation of item \(i\) in \(G^{R}_I\) \\
    \(h^{\star}_u\) & the graph enhance user preference of user \(u\) \\
    \(z^{\star}_i\) & the graph enhance user preference of item \(i\) \\
    \(m^i_u\) & the user interests of user \(u\) in candidate item \(i\) \\
    \(x^i_u\) & the integrated user interests of user \(u\) in candidate item \(i\) \\
    \(\alpha_{u,v}\) & the friends attention of friend \(v\) in contributing to \(x^i_u\) \\
    \(y^u_i\) & the item attraction of item \(i\) to targeted user \(u\) \\
    \(D\) & the number of embedding dimension \\
    \(W\) & the weight matrix in neural network \\
    \(\odot\) & the element-wise product operation \\
    \hline
\end{tabular}
}
\end{table}

\section{THE PROPOSED MODEL}

In this section, we give the overall architecture of the proposed model DICER and then introduce modules in detail. Finally, we discuss the training of the model.

\subsection{Architecture Overview}

\begin{figure*}[t]
    \centering
    \includegraphics[width=0.90\textwidth]{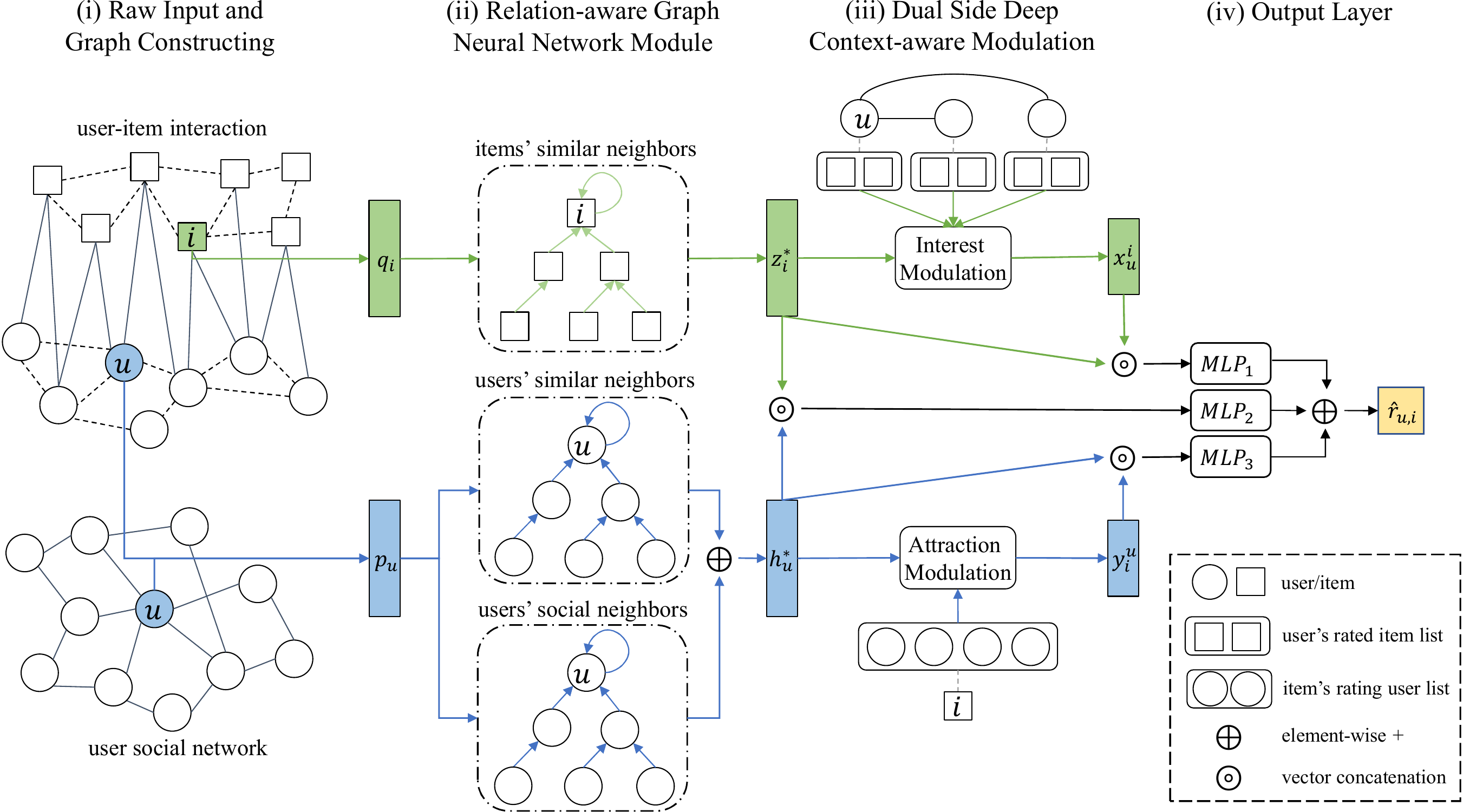}
	\caption{Architecture of the proposed DICER model with four modules. i) The Raw input which includes the user-item interaction network and user social network, we construct two collaborative similarity network based on the interaction network. ii) In the relation-aware GNN module, RGNN are to model the graph enhanced user preference \(h^{\star}_{u}\) and item attribute \(z^{\star}_i\). iii) In the dual side deep context-aware modulation layer, these graph enhanced representation is used as a deep context to capture user interest \(x^{i}_{u}\) and item attraction \(y^{u}_{i}\) from interaction history respectively. iv) Finally, user's preference and interest and item's attribute and attraction are input into the output layer to predict the final score \(\hat{r}_{u,i}\).}
	\label{fig:model}
\end{figure*}

The architecture of the proposed model is shown in Figure~\ref{fig:model}. The model consists of four modules. The first module is collaborative graph construction, which constructs two collaborative similarity graphs for users and items from user-item interactions, respectively. The second module is the high-order relation exploitation module, which is to learn the users' graph enhanced preference and items' graph enhanced attribute based on the social graph and the two collaborative similarity graphs. The third module is the dual side deep context-aware modulation, which is to model the user interest from user's and friends' interaction history based on the item's graph enhanced attribute and the item attraction from item's interaction history based on the user's graph enhanced preference. The fourth module is the user and item matching module based on the user's interest and preference and the item's attraction and attribute. We describe these four modules in detail next.

\subsection{Collaborative Graph Constructing}

\textbf{Collaborative Relation Definition}. 
We define \textit{collaborative similar users} as users who have similar consumption habits and \textit{collaborative similar items} as items with similar clicks or rated history. And one way to calculate the collaborative similarity is by the common interaction history of the users (items)~\cite{sarwar2001item}~\cite{wu2019dual}. For any user \(u\) and user \(v\), we define the strength of collaborative similarity \(sim_{u,v}\) between user \(u\) and user \(v\)  as below.
\begin{equation}
sim_{u,v} = \frac{ |R_I(u)\cap R_I(v)| }{ \sqrt{|R_I(u)|\cdot|R_I(v)|} }
\end{equation}
Then user \(u\) is collaborative similar to user \(v\) if and only if \( sim_{u,v}>\eta \) with \(\eta\) a fixed threshold, and the items' collaborative similarity relation is the same. 

We construct the collaborative similarity graph based on the collaborative similarity relation as \( G_U^{R} = (V_U, E_U^{R}) \) and \( G_I^{R} = (V_I, E_I^{R}) \) for users and items respectively, where \(V_U\) is the set of users and \(E_U^{R}\) is the set of edges that connect two collaborative similar users, and \(V_I\) is the set of items and \(E_I^{R}\) is the set of edges that connects two collaborative similar items.

\subsection{High-order Relation Exploitation}

\textbf{Relation-aware graph neural network module}. In order to exploit the high-order relation in social network \(G_U^{S}\) and collaborative similarity relation in \(G_U^{R}\) and \(G_I^{R}\), we propose a relation-aware Graph Neural Network (RGNN), an extension of NGCF which can efficient aggregate and select the most related information from neighbors. Let  \(z^{0}_i = \mathbf{q}_i\) and \(h^{S,0}_u = h^{R,0}_u = \mathbf{p}_u\)  as the first layer inputs for the RGNN module, by feeding the first layer inputs into three RGNN layers respectively, these layers can recursively model the propagation of item attribute or user preference in different graph. The detail process is as below:

\textit{I. RGNN to aggregate item-item collaborative similar neighbors}. For each item \(i\), given its \(l\)-th layer input \(z^{l}_i\), we model the updated user representation \(z^{l+1}_i\) at the \((l+1)\)-th layer from \(G^{R}_I\) as: 
\begin{equation}
\begin{aligned}
z^{l+1}_i &= AGG^R_I\left(z^{l}_i,\  z^{l}_j, \forall j \in N_I(i)\right) \\
&=\sigma\left( \sum_{j \in N_I(i)} \left(W^{I}_1 z^{l}_i + W^{I}_2( z^{l}_i \odot z^{l}_j)\right) \right)
\end{aligned}
\end{equation}
where  \(\sigma\) is the LeakyReLU activation function, \(\odot\) denotes the element-wise product, \(W^I_1, W^I_2\) are the weight matrix, \(N_I(i)\) is the collaborative similar neighbors of item \(i\) from \(G_I^{R}\) and \(z^{l}_j\) is the \(l\)-th layer representation of item \(j\). After the propagation process, the updated representation \(z^{l+1}_i\) aggregate the most related information from  the collaborative similar neighbors of \(l\)-th layer.

\textit{II. RGNN to aggregate social network neighbors}. For each user \(u\), given its \(l\)-th layer input \(h^{S,l}_u\), we model the updated user representation \(h^{S, l+1}_u\) at the \((l+1)\)-th layer from \(G^{S}_U\) as: 
\begin{equation}
\begin{aligned}
h^{S,l+1}_u &= AGG^S_U\left(h^{S,l}_u,\  h^{S, l}_f, \forall f \in F_U(u)\right) \\
&=\sigma\left( \sum_{f\in F_U(u)} \left(W^{S}_1 h^{S, l}_f + W^{S}_2( h^{S, l}_u \odot h^{S, l}_f)\right) \right)
\end{aligned}
\end{equation}
where \(W^{S}_1, W^{S}_2\) are the weight matrix, \(F_U(u)\) is the social neighbors of user \(u\) from \(G_U^{S}\) and \(h^{S, l}_f\) is the \(l\)-th layer representation of user \(f\). After the propagation process, the updated representation \(h^{S,l+1}_u\) aggregate the most related information from  the social network neighbors of \(l\)-th layer.

\textit{III. RGNN to aggregate user-user collaborative similar neighbors}. Similarly, for each user \(u\), given its \(l\)-th layer input \(h^{R,l}_u\), we model the updated user representation \(h^{R, l+1}_u\) at the \((l+1)\)-th layer from \(G^{R}_U\) as:
\begin{equation}
\begin{aligned}
h^{R,l+1}_u &= AGG^R_U\left(h^{R,l}_u,\  h^{R, l}_v, \forall v \in N_U(u)\right) \\
&=\sigma\left( \sum_{v\in N_U(u)} \left(W^{U}_1 h^{R, l}_f + W^{U}_2( h^{R, l}_u \odot h^{R, l}_v)\right) \right)
\end{aligned}
\end{equation}
where \(W^{U}_1, W^{U}_2\) are the weight matrix, \(N_U(u)\) is the collaborative similar neighbors of user \(u\) from \(G_U^{R}\) and \(h^{R, l}_v\) is the \(l\)-th layer representation of user \(v\).
After the propagation process, the updated representation \(h^{R,l+1}_u\) aggregate the most related information from the collaborative similar neighbors of \(l\)-th layer.

As users play a central role in both the social network \(G^{S}_U\) and collaborative similarity network \(G^{R}_U\), her final embedding \(h^{l}_u\) is a fusion of the aggregated neighbors information \(h^{S,l+1}_u\) and \(h^{R,l+1}_u\) from the \(G^{S}_U\) and \(G^{R_U}\) respectively. In fact, we try different kinds of fusion functions, including the concatenation and the addition, and find the addition always shows the best performance. Therefor, we use the addition as the fusion function in Eq.\eqref{user_add}
\begin{equation}
    h^{l}_u = h^{S,l}_u + h^{R,l}_u \label{user_add}
\end{equation}

Finally, After the iterative propagation process with \(l\) times in each RGNN layers, we obtain the updated representation set of \(u\) and \(i\) with \(h^{l}_u\) and \(z^{l}_i\) for \(l=[0,1,2,...,l]\). Then for each  user \(u\), her final graph enhanced preference is denoted as: \(h^{\star}_u = [h^{0}_u\|h^{1}_u\|...\|h^{l}_u]\) that concatenates her representation at each layer. Similarly, each item's final graph enhanced attribute is: \(z^{\star}_i = [z^{0}_i\|z^{1}_i\|...\|z^{l}_i]\). 

\subsection{Dual Side Deep Context-aware Modulation}

 Since only considering the candidate item or targeted user is unable to model the rich user interests or item attraction, in this section, we introduce our dual side deep context-aware modulation module which considers the graph enhanced user preference and item attribute as deep context to model the user interests and item attraction respectively. We will go into the details of each side modulation.

\subsubsection{\textbf{Deep context-aware user interest modulation}}
In social recommendation models, considering the information related to the candidate item from the user and friends' interaction history is necessary to model a better user interest~\cite{wu2019dual, fan2019deep}. However, consider the related information only based on the candidate item is not enough and may lead to missing information and a narrow understanding of user interest. In fact, users' interests in candidate's similar items are also useful.

In order to fully mine the information which can reflect the users' interest in the candidate item from their interaction history, we consider the graph enhanced item attribute as deep context which incorporates the high-order collaborative similar neighbors information from \(G^{R}_I\), then select the most related information from the user's rated items based on the deep context by a modulation operation as below:
 
\begin{equation}
m^{i}_{u} = f_I\left(z^{\star}_i,\ z^{\star}_j, \forall j \in R_I(u)\right)
\end{equation}
where \(R_I(u)\) is the rated items of user \(u\), and the \(f_I(\cdot)\) is:
\begin{equation}
m^{i}_{u} = MP_{j\in R_I(u)}(\{z^{\star}_{j,d} \odot z^{\star}_{i,d} \}), \forall d=1, ..., D
\end{equation}
where \( z^{\star}_{j,d}, z^{\star}_{i,d} \) are the \(d\)-th feature of \(z^{\star}_{j}, z^{\star}_{i}\) respectively. MP is the max pooling operation which can help to focus on the related information from the user's rated history based on the graph enhanced item attribute. Thus the obtained representation \(m^{i}_u\) capture the user \(u\)'s rich interest in candidate item \(i\) .

After obtained the user and her friend's interests in the candidate item, we aggregate the friend's interest with no-uniform weight. And the weight is varied when it comes to different candidate items. Intuitively, if a friend's interest in the candidate item is more similar to the targeted user, he should significantly influence the user's decision on the candidate item. Formally, given the user interest \(m^{i}_u\) and friend interest \(m^{i}_f\), we compute the attention weight \(\alpha_{u,f}\):
\begin{equation}
    \alpha_{u,f}^* = (m^{i}_u)^\top \cdot (m^{i}_f)
    \label{att_1}
\end{equation}
Then, the final attention weights of the friend interests are obtained by normalizing the
above attentive scores using softmax function, which can be interpreted as the influence of the friend to the final user interest representation of user \(u\) as:
\begin{equation}
\alpha_{u,f} = \frac{ exp(\alpha_{u,f}^*) }{ \sum_{f\in F_U(u)} exp(\alpha_{u,f}^*) }
\label{att_2}
\end{equation}
Finally, the final user interest (which integrated friends' interest) of user \(u\) is through the sum:
\begin{equation}
x^{i}_{u} = m^{i}_{u} + \sum_{f\in F_U(u)} \alpha_{u,f}\ m^{i}_{f}
\end{equation}
which considers both the users' own interest and the influence of her friends. Note that we tried a lot of strategies of combining the different features, such as concatenation, addition or full-connected neural layer, and find the addition always shows the best performance.

\subsubsection{\textbf{Deep context-aware item attraction modulation}} 
Similar to the user interest, the related information to the targeted user from item's interaction history can reflect the attraction of the candidate item~\cite{qin2020sequential, wu2019dual}. However, consider the related information only based on the targeted user is also not enough and may result in poor item attraction. In fact, if the candidate items' past consumers are the friends of the targeted user or have similar consumption habits to the targeted user, the item may have much attraction to the targeted user.

In order to fully mine the information which can reflect the items' attraction to the targeted user from their interaction history, we consider the graph enhanced user preference as deep context which incorporates the high-order collaborative similar neighbors information from \(G^{R}_U\) and soical neighbors information from \(G^S_U\), then select the most related information from the item's past consumers based on the deep context by a modulation operation as below:
\begin{equation}
y^{u}_{i} = f_U\left(h^{\star}_u,\ h^{\star}_v, \forall v \in R_U(i)\right)
\end{equation}
where \(R_U(i)\) is the user who rated item \(u\), and the \(f_U(\cdot)\) is:
\begin{equation}
y^{u}_{i} = MP_{v\in R_U(i)} (\{h^{\star}_{v,d} \odot h^{\star}_{u,d} \}), \forall d=1, ..., D
\end{equation}
where \(h^{\star}_{v,d}, h^{\star}_{u,d}\) are the \(d\)-th feature of \( h^{\star}_{v}, h^{\star}_{u}\) respectively. Thus the obtained representation \(y^{u}_i\) capture the item \(i\)'s rich attraction to targeted user \(u\) based on the user-user social relation and collaborative similarity relation.

\subsection{Output Layer}

Since user's decision on candidate item depend on both user preference and item attribute, we firstly predict the score based on the graph enhanced representation of user and item:
\begin{equation}
r^{O}_{ui} = MLP_1(h^{\star}_{u}, z^{\star}_{i})
\end{equation}
Next, the user’s interest in the item and the item’s attraction to the user reflect the matching score between them from two perspectives. Then we predict another two matching scores based on the two perspectives to make the model more robust.
\begin{equation}
r^{U}_{ui} = MLP_2(x^{i}_{u}, z^{\star}_{i})
\end{equation}
\begin{equation}
r^{I}_{ui} = MLP_3(y^{u}_{i}, h^{\star}_{u})
\end{equation}
The final predicted probability that user \(u\) will interact with item \(i\) is calculated by weighted-sum of the 3 scores, 
\begin{equation}
    \hat{r}_{u,i} = \lambda_1 r^{O}_{ui} + \lambda_2 r^{U}_{ui} + \lambda_3 r^{I}_{ui}
\end{equation}
where \(\lambda_1, \lambda_2\) and \(\lambda_3\) are hyperparameters, and \(\lambda_1+\lambda_2+\lambda_3=1,  \lambda_1 >= 0, \lambda_2 >= 0, \lambda_3 >= 0\).

\subsection{Model Training}

To learn model parameters of DICER, we need to specify an objective function to optimize. For implicit feed-back, the most widely adopted loss function is the cross-entropy defined as:
\begin{displaymath}
  \mathcal{L} = -\sum_{(u,i)} r_{u,i} \log \hat{r}_{u,i} + (1-r_{u,i}) \log (1-\hat{r}_{u,i}).
\end{displaymath}

To optimize the objective function, we adopt mini-batch Adaptive Moment Estimation (Adam)~\cite{kingma2015adam} as the optimizer in our implementation. Its main advantage is that the learning rate can be self-adapted during the training phase which eases the pain of choosing a proper learning rate. We also adopt the dropout strategy~\cite{srivastava2014dropout} to alleviate the overfitting issue in optimizing deep neural network models.

\section{EXPERIMENT}

To comprehensively evaluate the proposed model DICER~\footnote{The codes are released at \url{ https://github.com/Drone-Banks/DICER-WWW-2021}.}, we conduct experiments to answer the following research questions: \\
\textbf{RQ1} How does DICER perform compare with state-of-the-art methods for recommendation and social recommendation? \\
\textbf{RQ2} Are the key components in DICER, including relation-aware GNN module and deep context-aware modulation, necessary for improving performance? \\
\textbf{RQ3} How do hyper-parameters in DICER impact recommendation performance? \\

\subsection{Experiment Setup}

\subsubsection{\textbf{Datasets}} We conduct experiments on two representative datasets: \textit{Ciao} and \textit{Epinion}, which are taken from popular social network website Ciao~\footnote{\url{http://www.ciao.co.uk}} and Epinion~\footnote{\url{http://www.Epinion.com}}. Each social networking service allows users to clicked items and add friends. Hence, they provide a large amount of rating information and social information. 

\begin{table}
  \caption{Statistics of the datasets}
  \label{tab:datasets}
  \begin{tabular}{c|c|cl}
   \toprule
    Dataset & \textit{Ciao} & \textit{Epinion} \\
    \midrule
    Num. of Users & 7,375 & 20,608 \\
    \midrule
    Num. of Items & 106,797 & 23,585\\
    \midrule
    Num. of Ratings & 282,650 & 454,002 \\
    \midrule
    Num. of Relations & 111,781 & 351,486 \\
    \midrule
    Rating Density & 0.0359\% & 0.0934\%\\
    \midrule
    Relation Density & 0.2055\% & 0.0828\%\\
  \bottomrule
\end{tabular}
\end{table}

As long as some user-user or user-item interactions exist, the corresponding rating is assigned a value of 1 as implicit feedback. The statistical details of these datasets are summarized in Table \ref{tab:datasets}

\subsubsection{\textbf{Baselines}} To illustrate the effectiveness of our model, we compare DICER with two classical collaborative filtering (CF) models, two social-based recommendation models, two deep learning based recommendation models, and two deep-learning based social recommendation models.

The first group of models are CF models:

\begin{itemize}
\item {\textbf{BPR}}~\cite{rendle2009bpr}: This is a competing latent factor model for implicit feedback based recommendation. It designed a ranking based function that assumes users prefer items they like compared to unobserved ones. 
\item {\textbf{FM}}~\cite{rendle2010factorization}: This model is a unified latent factor based model that leverages the user and item attributes. In practice, we use the user and item features as introduced above.
\end{itemize}

The second group contains social based recommendation models that utilize social relation information:

\begin{itemize}
\item {\textbf{TrustMF}}~\cite{yang2017social}: This is a matrix factorization method, which maps users into two spaces and opti mizes user embedding to retrieve the trust matrix.
\item {\textbf{TrustSVD}}~\cite{guo2015trustsvd}:This is another matrix factorization-based method, wich incorporates friends’ embedding vectors into targeted user’s predicted rating.
\end{itemize}

The third group is deep learning based recommendation models:

\begin{itemize}

\item {\textbf{NCF}}~\cite{he2017neural}: This is a  deep learning based recommendation model which leverage a multi-layer perceptron to learn the user-item interaction function.
\item {\textbf{NGCF}}~\cite{wang2019neural}: This is a graph based recommendation model that model the high-order connectivity in the user-item graph and inject the collaborative signal into the embedding process in an explicit manner.
\end{itemize}

The fourth group is deep-learning based social recommendation models:
\begin{itemize}
\item {\textbf{SAMN}}~\cite{chen2019social}: This is a strong baseline for social recommendation, which leverages attention mechanisms to model both aspect- and friend-level differences for social-aware recommendation.
\item {\textbf{DiffNet++}}~\cite{wu2020diffnetplus}: This is another strong baseline, which adopts GNN and considers both the social influence diffusion and the latent collaborative interests diffusion in social-aware recommendation.
\end{itemize}

\begin{table}
  \caption{Comparison of the methods. For social domain, we use "S" represent the social information and "HS" represent the high-order social information. For item domain, we use "I" represent the interest information and "HI" represent the high-order interest information. For user interest and item attraction, we use "SC" denotes shallow context-aware and "DC" denotes deep context-aware. And we use "DL" denote deep learning based methods.}
  \label{tab:camparison_methods}
  \resizebox{\linewidth}{!}{
  \begin{tabular}{c|c|c|c|c|c|c|c|c|c}
  \toprule
    \multirow{2}{*}{Models} & \multicolumn{2}{c|}{Social Domain} & \multicolumn{2}{c|}{Item Domain} & \multicolumn{2}{c|}{User Interest} & \multicolumn{2}{c|}{Item Attraction} & \multirow{2}{*}{DL} \\  
    \cline{2-9}
     & S & HS & I & HI & SC & DC & SC &DC &\\

    \hline
    TrustMF & \(\surd\) & \(\backslash\) & \(\surd\) & \(\backslash\) & \(\backslash\) & \(\backslash\) & \(\backslash\) & \(\backslash\) & \(\backslash\) \\
    \hline
    TrustSVD& \(\surd\) & \(\backslash\) & \(\surd\) & \(\backslash\) & \(\surd\) & \(\backslash\) & \(\backslash\) & \(\backslash\) & \(\backslash\) \\
    \hline
    NCF     & \(\backslash\) & \(\backslash\) & \(\surd\) & \(\backslash\) & \(\backslash\) & \(\backslash\) & \(\backslash\) & \(\backslash\) & \(\surd\) \\
    \hline
    NGCF    & \(\backslash\) & \(\backslash\) & \(\surd\) & \(\surd\) & \(\backslash\) & \(\backslash\) & \(\backslash\) & \(\backslash\) & \(\surd\) \\
    \hline
    SAMN    & \(\surd\) & \(\backslash\) & \(\surd\) & \(\backslash\) & \(\surd\) & \(\backslash\) &  \(\backslash\) & \(\backslash\) & \(\surd\) \\
    \hline
    DiffNet++&\(\surd\) & \(\surd\) & \(\surd\) & \(\surd\) & \(\backslash\) & \(\backslash\) &  \(\backslash\) & \(\backslash\) & \(\surd\) \\
    \hline
    DICER   & \(\surd\) & \(\surd\) & \(\surd\) & \(\surd\) & \(\surd\) & \(\surd\) & \(\surd\) & \(\surd\) & \(\surd\) \\
  \bottomrule
\end{tabular}}
\end{table}

The comparison of DICER and the baseline methods are listed in Table \ref{tab:camparison_methods}.
\begin{table*}[!tbp]
  \caption{Comparisons of different methods on Two datasets. Best baselines are underlined. The proposed method achieves best performances on all metrics which are in boldface. The last column “RI” indicates the relative improvement of DICER over the corresponding baseline on average.}
  \label{tab:main_comparision}
  \begin{tabular}{c|c|c|c|c|c|c|cl}
    \hline
    $\bm{Ciao}$ & $\textbf{Recall@5}$ & $\textbf{Recall@10}$ & $\textbf{Recall@15}$ & $\textbf{NDCG@5}$ & $\textbf{NDCG@10}$ & $\textbf{NDCG@15}$ & $\textbf{RI}$ \\
    \hline
    \textbf{BPR} & 0.1782 & 0.2143 & 0.2469 & 0.1618 & 0.1720 & 0.1814 & +42.84\% \\
    \textbf{FM} & 0.1852 & 0.2269 & 0.2613 & 0.1638 & 0.1760 & 0.1861 &  +37.83\% \\
    \hline
    \textbf{TrustMF} & 0.2151 & 0.2631 & 0.3027 & 0.1916 & 0.2062 & 0.2179& +18.28\%\\
    \textbf{TrustSVD} & 0.2159 & 0.2698 & 0.3117 & 0.1884 & 0.2056 & 0.2179 &  +17.53\%\\
    \hline
    \textbf{NCF} & 0.1840 & 0.2268 & 0.2609 & 0.1644 & 0.1773 & 0.1873 & +37.62\%\\
    \textbf{NGCF} & 0.2330 & 0.2821 & 0.3185 & 0.2063 & 0.2212 & 0.2319 &  +10.53\% \\
    \hline
    \textbf{SAMN} & 0.2322 & 0.2836 & 0.3245 & 0.2030 & 0.2205 & 0.2332 &  +10.40\% \\
    \textbf{DiffNet++} & \underline{0.2330} & \underline{0.2844} & \underline{0.3259} & \underline{0.2063} & \underline{0.2226} & \underline{0.2351} &  +9.59\% \\
    \hline
    \textbf{DICER} & \textbf{0.2554} & \textbf{0.3151} & \textbf{0.3579} & \textbf{0.2243} & \textbf{0.2437} & \textbf{0.2565} & - \\
    \hline
  \end{tabular}
  
  \begin{tabular}{c|c|c|c|c|c|c|cl}
    \hline
    $\bm{Epinion}$ & $\textbf{Recall@5}$ & $\textbf{Recall@10}$ & $\textbf{Recall@15}$ & $\textbf{NDCG@5}$ & $\textbf{NDCG@10}$ & $\textbf{NDCG@15}$ & $\textbf{RI}$\\
    \hline
    \textbf{BPR} & 0.1616 & 0.2264 & 0.2716 & 0.1253 & 0.1484 & 0.1622 & +44.06\% \\
    \textbf{FM} & 0.1592 & 0.2273 & 0.2763 & 0.1233 & 0.1476 & 0.1627 &  +44.38\% \\
    \hline
    \textbf{TrustMF} & 0.1816 & 0.2602 & 0.3163 & 0.1374 & 0.1651 & 0.1821& +27.75\%\\
    \textbf{TrustSVD} & 0.1927 & 0.2623 & 0.3090 & 0.1466 & 0.1712 & 0.1852 & +24.31\% \\
    \hline
    \textbf{NCF} & 0.1834 & 0.2624 & 0.3187 & 0.1397 & 0.1675 & 0.1844 & +26.28\%\\
    \textbf{NGCF} & 0.2099 & 0.2918 & 0.3488 & 0.1618 & 0.1908 & 0.2080 &  +11.80\% \\
    \hline
    \textbf{SAMN} & 0.2206 & 0.3055 & 0.3625 & 0.1697 & 0.1996 & 0.2170 &  +6.89\% \\
    \textbf{DiffNet++} & \underline{0.2298} & \underline{0.3183} & \underline{0.3786} & \underline{0.1742} & \underline{0.2055} & \underline{0.2236} &  +3.21\% \\
    \hline
    \textbf{DICER} & \textbf{0.2370} & \textbf{0.3269} & \textbf{0.3854} & \textbf{0.1818} & \textbf{0.2134} & \textbf{0.2312} & - \\
    \hline
  \end{tabular}
\end{table*}

\subsubsection{\textbf{Evaluation Metrics}} We adopt \textit{Recall@K} and \textit{NDCG@K} to evaluate the performance of all methods. The two metrics have been widely used in previous recommendation studies~\cite{chen2019social, xiao2017learning, yu2018aesthetic}. \textit{Recall@K} considers whether the ground truth is ranked among the top \(K\) items, while \textit{NDCG@K} is a position-aware ranking metric.

\subsubsection{\textbf{Experiments Details}} The proposed DICER was implemented with PyTorch and we use the Xavier initializer~\cite{glorot2010understanding} to initialize the model parameters. For each dataset, we use 80\% as a training set to learn parameters, 10\% as a validation set to tune hyper-parameters and 10\% as a test set for the final performance comparison. The hyper-parameter settings are as follow: learning rate is 0.001, training batch size is 4096, embedding size \(D=64\), the RGNN layers \(l=3\), coefficient \(\lambda_1=\lambda_2=\lambda_3=\frac{1}{3}\), collaborative similarity threshold \(\eta = 0.1\), the LeakyReLU slope is 0.2. In the training process, as there are many more unobserved items for each user, we randomly select 8 times pseudo negative samples for each user at each iteration. Since each iteration we change the pseudo negative samples, each unobserved item gives a weak signal. For all the baselines, we carefully tune the parameters to ensure the best performance.

\subsection{Comparative Results: RQ1}

The comparison of different methods on two datasets is shown in Table  \ref{tab:main_comparision}. We set the length K = 5, 10, and 15 in our experiments to evaluate on different recommendation lengths. From the results, the following observations can be made:


First, methods incorporating social information generally perform better than non-social methods. For example, in Table \ref{tab:main_comparision}, the performance of TrustSVD and TrustMF is better than BPR and FM, and SAMN, DiffNet++ and DICER outperform NGCF and NCF. This is consistent with previous work~\cite{chen2019social, xiao2017learning, zhao2014leveraging}, which indicates that social information is helpful to improve the recommendation performance.

Second, our method DICER achieves the best performance on the two datasets. Specifically, compared to SAMN –  an attention-based deep learning model, DICER exhibits average improvements of 10.40\% and 6.89\% on the two datasets. And compared to DiffNet++ - a recently proposed and very expressive GNN-based model, DICER exhibits average improvements of 9.59\% and 3.21\% on the two datasets. The substantial improvement of our model over the baselines could be attributed to two reasons: (1) our model use relation-aware GNN to deal with the high-order social relation and collaborative similarity relation, which allow the related information from multi-relation neighbors to be utilized; (2) we model the users' interests and items' attraction based on the deep context, i.e., the graph enhanced user and item representation.

\subsection{Study of DICER: RQ2}

In this subsection, we study the impact of model components.

\subsubsection{Effect of Dual Side information} In the last subsection, we have demonstrated the effectiveness of the proposed model. The models provide dual side modulation to (1) capture the more diverse user interest and (2) model the rich item attraction. To prove the effective of the dual side information, we compare DICER with its two variants: DICER w/o ui, and DICER w/o ia. These two variants are defined in the following:
\begin{itemize}
    \item DICER w/o ui: The user side modulation and the user interest information is removed. This variant only use the users' graph enhanced preference \(\mathbf{h}^{\star}_u\) and items' graph enhanced attribute \(\mathbf{z}^{\star}_i\) and items' attraction \(y^{u}_i\) to predict the score, while ignoring the user interest \(x^{i}_u\).
    \item DICER w/o ia: The item side modulation and the item attraction information is removed. This variant only use the items' graph enhanced attribute \(\mathbf{z}^{\star}_i\) and users' graph enhanced preference \(\mathbf{h}^{\star}_u\) and users' interest \(x^{i}_u\) to predict the score, while ignoring the item attraction \(y^{u}_i\).
\end{itemize}

\begin{figure*}[tbp]
    \centering
    \includegraphics[width=\textwidth]{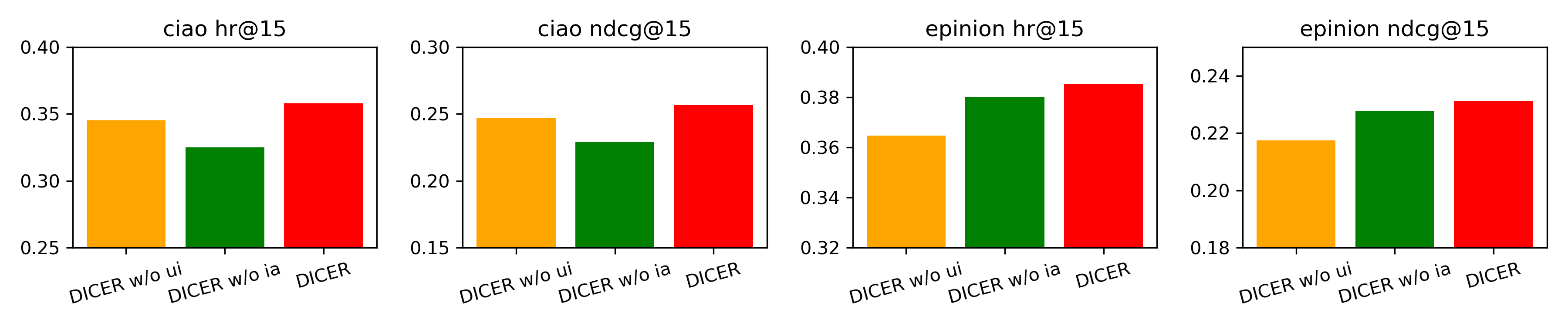}
	\caption{Effect of dual side information on Ciao and Epinion datasets}
	\label{fig:dual_side}
\end{figure*}

The performance of DICER and its variants on Ciao and Epinion are given in Figure \ref{fig:dual_side}. From the results, we have the following findings:
\begin{itemize}
    \item \textbf{User Interest}: We now focus on analyzing the effectiveness of user interest, which combine the related information extracted from the user and friends' interaction history. DICER w/o ui performs worse than DICER, On average, the relative reduction is 3.53\% on NDCG metric and 4.54\% on Recall metric. It verifies that the user interest information is important to predict the score and boost the recommendation performance.
    \item \textbf{Item Attraction}: We can see that without the item attraction, the performance decrease significantly. On average, the relative reduction is 5.14\% on NDCG metric and 5.79\% on Recall metric, respectively. It justifies our assumption that item attraction has informative information that can help to predict the final score and improve the performance of recommendation. 
\end{itemize}

\begin{table}[tbp]
  \caption{Effect of deep context and  modulation on Ciao}
  \label{tab:context_modulation}
  \resizebox{\linewidth}{!}{
  \begin{tabular}{c|ccc|cccl}
  \toprule
    Models & recall@5 & recall@10 & recall@15 & ndcg@5 & ndcg@10 & ndcg@15 \\
    \midrule
    DICER-\(\alpha\) & 0.2307 & 0.2842 & 0.3302 & 0.2030 & 0.2193 & 0.2327 \\
    DICER-\(\beta\) & 0.2340 & 0.2928 & \underline{0.3380} & 0.2066 & 0.2251 & 0.2388 \\
    DICER-\(\mu\) & \underline{0.2401} & \underline{0.2945} & 0.3357 & \underline{0.2108} & \underline{0.2289} & \underline{0.2401} \\
    DICER-\(\alpha \& \beta \& \mu\) & 0.2187 & 0.2733 & 0.3169 & 0.1918 & 0.2093 & 0.2222 \\
    \midrule
    DICER-\(attn\) & 0.2150 & 0.2688 & 0.3097 & 0.1875 & 0.2054 & 0.2179 \\
    \midrule
    DICER & \textbf{0.2554} & \textbf{0.3151} & \textbf{0.3579} & \textbf{0.2243} & \textbf{0.2437} & \textbf{0.2565} & \\
  \bottomrule
\end{tabular}}
\end{table}

\subsubsection{Effect of deep context and modulation} To get a better understanding of the proposed DICER model, we further do some ablation studies for the key components of DICER - the deep context-aware modulation and the results are shown in Table \ref{tab:context_modulation}. There are three different relations for deep context, including item's collaborative similarity \(\alpha\), user's collaborative similarity \(\beta\), user's social relation \(\mu\). We first compare DICER with its four variants: DICER-\(\alpha\), DICER-\(\beta\), DICER-\(\mu\) and DICER-\(\alpha \& \beta \& \mu\). Moreover, we use the attention mechanism to replace the modulation as the variant DICER-\(attn\) to prove the effectiveness of modulation. These five variants are defined in the following:
\begin{itemize}
    \item DICER-\(\alpha\): The item's collaborative similarity relation is eliminated during modeling the user interest. This variant only considers the candidate item as the context.
    \item DICER-\(\beta\): The user's collaborative similarity relation is eliminated during modeling the item attraction. This variant only considers the targeted user and her social neighbors as the context.
    \item DICER-\(\mu\): The user's social relation is eliminated during modeling the item attraction. This variant only considers the targeted user and her collaborative similar neighbors as the context.
    \item DICER-\(\alpha \& \beta \& \mu\): This variant eliminates the three relations (item's collaborative similarity \(\alpha\), user's collaborative similarity \(\beta\), user's social relation \(\mu\)) and onyl consider the candidate item or targted user as context.
    \item DICER-\(attn\): This variant replaces the modulation function with a attention mechanism like Eq.\eqref{att_1} and Eq.\eqref{att_2} to select the related information from interaction history.
\end{itemize}

The results of different relation based context on DICER and the DICER-\(attn\) are shown in Table \ref{tab:context_modulation}. From the results, we have the following findings:
\begin{itemize}
    \item The three relations all contribute to the deep context-aware modulation for modeling the user interest and item attraction. And the more relation the deep context is based on, the context-aware information can be better for improving the performance.
    \item The modulation function is more effective than the attention mechanism when model the related information from the interaction history, This may due to the max-pooling operation in modulation is aiming to select the most similar feature which is more effect than the attention mechanism.
\end{itemize}

\begin{table}[tbp]
  \caption{Effect of GNN module on Ciao}
  \label{tab:gnn_module}
  \resizebox{\linewidth}{!}{
  \begin{tabular}{c|ccc|cccl}
  \toprule
    Models & recall@5 & recall@10 & recall@15 & ndcg@5 & ndcg@10 & ndcg@15 \\
    \midrule
    DICER-embed & 0.1823 & 0.2190 & 0.2535 & 0.1610 & 0.1718 & 0.1818 \\
    DICER-GAT & 0.2498 & 0.3077 & 0.3548 & 0.2182 & 0.2369 & 0.2511 \\
    DICER-RGNN & \textbf{0.2554} & \textbf{0.3151} & \textbf{0.3579} & \textbf{0.2243} & \textbf{0.2437} & \textbf{0.2565} & \\
  \bottomrule
\end{tabular}}
\end{table}

\subsubsection{Effect of GNN module} To prove the effectiveness of the RGNN module, We denote our model as DICER-RGNN and simply our model as DICER-embed and DICER-GAT, while DICER-embed removes the GNN module and DICER-GAT replaces the GNN module with GAT. And we compare the three variants with DICER. As we can see in Table~\ref{tab:gnn_module}, GAT improve performance by specifying different weight to aggregate the neighbors' information. Moreover, we can find that DICER with the RGNN module further improves performance by aggregate more related information via the max-pooling operation in modulation (about extra 2.6\% impv. for NDCG over DICER-GAT). The result show that the most related information from the multi-relation neighbors can bring significant performance improvement.

To sum up, DICER can leverage the high-order multi-relation information and model the rich user interests and item attraction via the deep context-aware modulation, which can boost the recommendation performance.

\begin{figure*}[tbp]
    \centering
    \includegraphics[width=\textwidth]{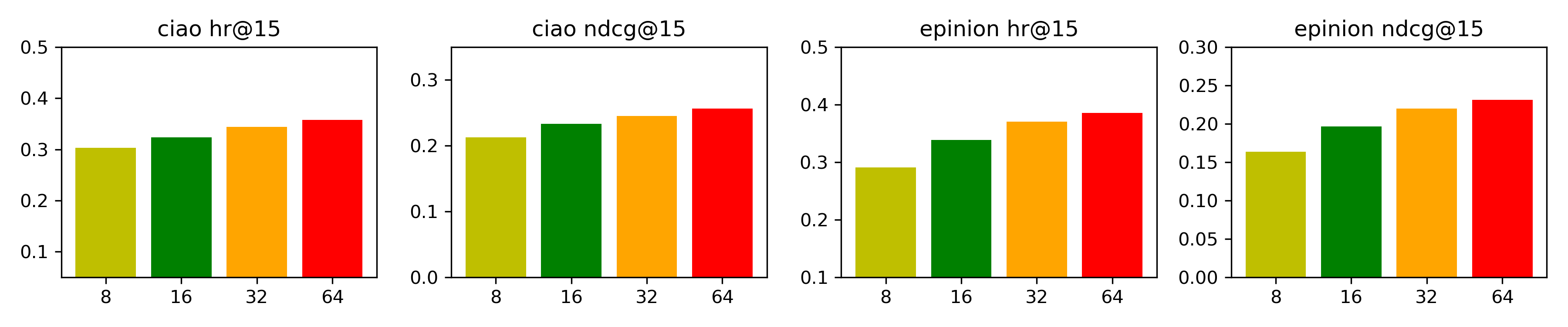}
	\caption{Effect of embedding size \(D\) on Ciao and Epinion datasets. Noted that, the graph enhanced embedding size will be \(4\times D\).}
	\label{fig:embed_size}
\end{figure*}

\subsection{Parameter Sensitivity: RQ3} 

In this subsection, we first analyze the effect of embedding size of user embedding \(P\), item embedding \(Q\), on the performance of our model. It should be noted that, if the embedding size is \(D\), the graph enhanced embedding vector's size will be \((l+1) \times D\) and it is \(4D\) in our implementation. Figure \ref{fig:embed_size} presents the performance comparison w.r.t. the length of embedding of our proposed model on Ciao and Epinion datasets. In general, with the increase of the embedding size, the performance also increases. As we can see, increasing the embedding size from 8 to 32 can bring significant performance improvement, a further increase to 64 brings less improvement but a larger computational complexity. Therefore,  we need to choose an appropriate embedding size to balance model performance and efficiency.

\begin{table}
  \caption{Effect of rgnn layer numbers on Ciao}
  \label{tab:gnn_layers}
  \resizebox{\linewidth}{!}{
  \begin{tabular}{c|ccc|cccl}
  \toprule
    Models & recall@5 & recall@10 & recall@15 & ndcg@5 & ndcg@10 & ndcg@15 \\
    \midrule
    DICER-1 & 0.2526 & 0.3074 & 0.3534 &  0.2211 & 0.2385 & 0.2521 \\
    DICER-2 & 0.2530 & 0.3088 & 0.3530 & 0.2226 & 0.2404 & 0.2537 \\
    DICER-3 & \textbf{0.2554} & \textbf{0.3151} & \textbf{0.3579} & \textbf{0.2243} & \textbf{0.2437} & \textbf{0.2565} & \\
  \bottomrule
\end{tabular}}
\end{table}

Secondly, to investigate whether DICER can benefit from multiple RGNN propagation layers, we search the layer numbers in the range of \{1,2,3\}. Table \ref{tab:gnn_layers} summarizes the experimental results, where DICER-\(n\) indicates the model with \(n\) RGNN propagation layers. As we can see, increasing the number of propagation layers substantially enhances the recommendation performance. We attribute the improvement to the effective modeling of neighbors information: the related information from neighbors can enhance the user preference and item attribute, which contribute to the deep context-aware modulation, respectively.

\section{RELATED WORK}

In this section, we briefly review the related work about the social recommendation, graph neural network techniques employed for recommendation, and the context-aware recommendation.

\textbf{Social Recommendation}. In recent years, there are lots of works exploiting user's social relations for improving the recommender system~\cite{wu2021collaborative, yang2017social, tang2013exploiting, tang2016recommendations}. Most of them assume that users' preference is similar to or influenced by their friends, which can be suggested by social theories such as social homophily~\cite{mcpherson2001birds} and social influence~\cite{marsden1993network}. According to the assumptions above, social regularization has been proposed to restrain the user embedding learning process in the latent factor based models~\cite{hu2015synthetic, ma2011recommender, jamali2010matrix, ma2008sorec}. And TrustMF~\cite{yang2017social} model is proposed to model the mutual influence between users by mapping users into two low-dimensional space: truster space and trustee space and factorize the social trust matrix. By treating the social neighbors' opinion as the auxiliary implicit feedbacks of the targeted user, TrustSVD~\cite{guo2015trustsvd} is proposed to incorporate the social influence from social neighbors on top of SVD++~\cite{koren2008factorization}.  Moreover, some recent studies like~\cite{wang2017item, fan2018deep, chen2019social} and~\cite{fan2019deep, chen2019an, krishnan2019a} leverage deep neural network and transfer learning or adversarial learning approach respectively, to learn a more complex representation or model the shared knowledge between social domain and item domain. However, comparing with our models in this paper, the common limitations of existing studies are: i) they did not leverage the high-order social relation and collaborative relation among users; ii) they ignore the related information from interaction history based on the relation enhanced deep context.

\textbf{Graph Neural Network for Recommendation}. More recently, Graph Neural Networks (GNNs) have been proven to have great potential to learn the graph structure data~\cite{NIPS2016_6081, kipf2017semi, derr2018signed}. 
In the task of recommender systems, the user-item interaction records obviously form a typical graph. Hence, there are many works that adopted GNNs to solve the recommendation problem~\cite{berg2017graph, ying2018graph, wang2019neural}. GCMC~\cite{berg2017graph} proposed a graph auto-encoder framework to predict unobserved interactions in the user-item matrix. Pinsage~\cite{ying2018graph} proposed a random-walk graph neural network to learn the node embeddings in web-scale graphs. And NGCF~\cite{wang2019neural} proposed a multi-layer GNNs which can model the higher-order collaborative signals between users and items during the users and items embedding learning process. As the social relation among users could be naturally formulated as a user-user graph, there are also some works~\cite{fan2019graph, wu2019a,wu2019dual, wu2020diffnetplus} using GNNs to capture social information for recommendation. The Diffnet++~\cite{wu2020diffnetplus} developed a GNN based model to simulate both the social influence and user interest diffusion process. And DANSER~\cite{wu2019dual} developed a dual graph attention networks to model the two-fold social effects collaboratively. Although these works model the social influence with GNNs, they didn't take advantage of high-order relations. Our work differs from these works. We model the high-order relation with multi-relation GNNs and consider the relation enhanced information as a deep context to extract the related information from interaction history.

\textbf{Context-aware Recommendation}. Context-aware recommendation aim to further improve performance accuracy and user satisfaction by fully utilizing the contextual information~\cite{li2012context}. And the basic context is the candidate item and targeted user. Such as~\cite{zhou2018deep} proposed a deep interest network to adaptively learn the user interests from historical behaviors concerning a candidate item. As for the social recommendation,~\cite{fan2019deep} proposed DSCF to consider the local and distant social neighbors' information under the specific recommendation context. And SAMN~\cite{chen2019social} proposed an attention-based memory module to extract the most related information from the social neighbors. Compared to these methods, our model possesses two key difference: i) For modeling the user's and friends' interest from the interaction history, we further consider the collaborative relations among items as a deep context ; ii) For modeling the item' attraction to the targeted user, we are the first to consider the social relation and collaborative relation among users as a deep social context and capture the more related information from item's past consumers.

\section{CONCLUSION}

In this paper, we proposed DICER, which utilizes a multi-relation graph neural network to learn the graph information and extract the most related information under the graph enhanced deep context. Our method is equipped with good performance because: i) The multi-relation graph neural network module can capture the high-order relation information in both social graph and collaborative similarity graph. ii) The dual side deep context-aware modulation can model the rich user interest and item attraction from the interaction history. Our comparative experiments and ablation studies on the two benchmark datasets showed that the multi-relation graph neural network module could model the better high-order relation information. The deep context-aware modulation plays a crucial role in both the user and item side.

For future work, we plan to investigate the dual side information fusion strategy further. Moreover, we also seek to deploy our method on real-world recommender systems.

\section*{Acknowledgments}

We want to express gratitude to the anonymous reviewers for their hard work. This work is funded by NSFC 61976114 and NSFC 61936012. This work is also partially supported by the research funding from ZTE Corporation.





\bibliographystyle{ACM-Reference-Format}
\balance
\bibliography{reference}

\end{document}